\documentclass[1p,times]{elsarticle}

\usepackage{amssymb}
\usepackage[numbers]{natbib}
\usepackage{epstopdf}
\usepackage{color}
\usepackage{booktabs}
\usepackage{tabularx}
\usepackage{enumitem}
\usepackage{graphics}
\usepackage{graphicx}
\usepackage{float}
\usepackage{xcolor}
\usepackage{subfigure}
\usepackage{subfig}
\usepackage{mathtools}
\usepackage{amsmath}
\usepackage{array}
\usepackage{longtable}
\usepackage[noend]{algpseudocode}
\usepackage{algorithmicx,algorithm}
\usepackage{multirow}
\usepackage{diagbox}
\usepackage{stfloats}
\usepackage{ragged2e}
\usepackage{makecell}
\begin{document}

\begin{frontmatter}

%% Title, authors and addresses

%% use the tnoteref command within \title for footnotes;
%% use the tnotetext command for theassociated footnote;
%% use the fnref command within \author or \address for footnotes;
%% use the fntext command for theassociated footnote;
%% use the corref command within \author for corresponding author footnotes;
%% use the cortext command for theassociated footnote;
%% use the ead command for the email address,
%% and the form \ead[url] for the home page:
%% \title{Title\tnoteref{label1}}
%% \tnotetext[label1]{}
%% \author{Name\corref{cor1}\fnref{label2}}
%% \ead{email address}
%% \ead[url]{home page}
%% \fntext[label2]{}
%% \cortext[cor1]{}
%% \affiliation{organization={},
%%             addressline={},
%%             city={},
%%             postcode={},
%%             state={},
%%             country={}}
%% \fntext[label3]{}

\title{Research and application of artificial intelligence based webshell detection model: A literature review\tnoteref{t1}}
\tnotetext[t1]{This paper is supported by the National Key Research and Development Program of China: 2022YFB3103403, the Fundamental Research Funds for the Central Universities: YCJJ20230464, and the National Natural Science Foundation of China: 6217071437,62072200,62127808}

%% use optional labels to link authors explicitly to addresses:
%% \author[label1,label2]{}
%% \affiliation[label1]{organization={},
%%             addressline={},
%%             city={},
%%             postcode={},
%%             state={},
%%             country={}}
%%
%% \affiliation[label2]{organization={},
%%             addressline={},
%%             city={},
%%             postcode={},
%%             state={},
%%             country={}}

\author[a]{Mingrui Ma\fnref{fn1}}
\ead{m202271767@hust.edu.cn}
\author[a]{Lansheng Han\corref{cor1}%
\fnref{fn2}}
\ead{1998010309@hust.edu.cn}
\author[b]{Chunjie Zhou\fnref{fn3}}
\ead{cjiezhou@hust.edu.cn}

\fntext[fn1]{His research interests include information security, network security, neural network, deep learning and artificial intelligence}
\fntext[fn2]{He has published more than 50 papers in various well-known journals and conferences}
\fntext[fn3]{His research interests include safety and security control of industrial control systems, theory and networked control systems, and artificial intelligence applications of Industrial Internet system}

\cortext[cor1]{Corresponding author}
\address[a]{Hubei Key Laboratory of Distributed System Security, Hubei
Engineering Research Center on Big Data Security, School of Cyber Science and
Engineering, Huazhong University of Science and Technology, Wuhan, 430074, China}
\address[b]{The Key Laboratory of Ministry of Education for Image Processing and Intelligent Control, School of Artificial Intelligence and Automation, Huazhong University of Science and Technology, Wuhan, 430074, Hubei, China}

\begin{abstract}
%% Text of abstract
Webshell, as the "culprit" behind numerous network attacks, is one of the research hotspots in the field of cybersecurity. However, the complexity, stealthiness, and confusing nature of webshells pose significant challenges to the corresponding detection schemes. With the rise of Artificial Intelligence (AI) technology, researchers have started to apply different intelligent algorithms and neural network architectures to the task of webshell detection. However, the related research still lacks a systematic and standardized methodological process, which is confusing and redundant. Therefore, following the development timeline, we carefully summarize the progress of relevant research in this field, dividing it into three stages: Start Stage, Initial Development Stage, and In-depth Development Stage. We further elaborate on the main characteristics and core algorithms of each stage. In addition, we analyze the pain points and challenges that still exist in this field and predict the future development trend of this field from our point of view. To the best of our knowledge, this is the first review that details the research related to AI-based webshell detection. It is also hoped that this paper can provide detailed technical information for more researchers interested in AI-based webshell detection tasks.
\end{abstract}

%%Graphical abstract
%\begin{graphicalabstract}
%\includegraphics{grabs}
%\end{graphicalabstract}

%%Research highlights
%%\begin{highlights}
%%\item Research highlight 1
%%\item Research highlight 2
%%\end{highlights}

\begin{keyword}
%% keywords here, in the form: keyword \sep keyword
Artificial intelligence \sep Webshell detection \sep Machine learning \sep Neural network \sep Language model
%% PACS codes here, in the form: \PACS code \sep code

%% MSC codes here, in the form: \MSC code \sep code
%% or \MSC[2008] code \sep code (2000 is the default)

\end{keyword}

\end{frontmatter}

%% \linenumbers

%% main text
\section{Introduction} \label{1}
In the context of big data, cloud computing, and the Internet+, web servers have gradually become hot targets for cyber attacks. According to the Open Web Application Security Project (OWASP), injection vulnerability has become one of the top ten vulnerabilities. Attackers can inject or directly upload malicious scripts or attack programs to web servers through files to perform unauthorized operations. (i.e. remote access control, privilege escalation, access to sensitive data, etc.) Webshell, as one of the typical representatives of malicious scripts, has a variety of characteristics. It can be a single line of code (i.e. one-sentence Trojan) that allows remote execution of user-provided system commands, or it can be a complex script file consisting of a huge amount of code. Webshell also has a variety of forms, including common file formats (i.e. ASP, ASPX, PHP, JSP, PL, PY, etc.) and even high-resolution images. In addition, attackers employ techniques such as inserting unrelated code, code obfuscation, program packing, function hiding, string encoding, etc. to hide and disguise webshells, bypassing the rule-based and signature-based matching tools. Webshells are difficult to leave a complete record in system logs, making it difficult for system administrators to trace back to their source. 

Given the various circumstances mentioned above, accurate detection of webshell is not an easy task. The key to solving this problem lies in identifying and distinguishing the feature differences between webshell-related data and normal data. Abdelhakim et al. \cite{RF1} classified webshell features into five categories: lexical features, syntactical features, semantical features, static features, and abstract features. Among them, abstract features are defined as advanced features that go beyond syntactical, semantical, and lexical features, which help to reveal hidden parts of webshell that cannot be detected by syntactical and semantical analysis. Specifically, abstract features are vectorized data that can represent source code, opcode, network traffic, etc., making them primarily suitable for neural network and deep learning methods. Abdelhakim et al. \cite{RF1} have demonstrated that AI techniques have been able to achieve better detection performance compared to static or rule-based methods. However, their work merely provides a brief summary and overview of related research without delving into a detailed exploration and analysis of the methods. Therefore, this paper focuses on summarizing webshell detection methods based on AI techniques, analyzing their connections and differences, pointing out the defects and challenges of the methods, and providing insights into future development trends.

The rest of the paper is organized as follows:

Section \ref{2} classifies and introduces the relevant research in terms of development timeline, model types (e.g. machine learning, deep learning, and hybrid models), and detection datatypes (i.e. source code, opcode/bytecode, traffic data/flow data, etc); Section \ref{3} summarizes and analyzes in detail the key issues and challenges faced by relevant researches; Section \ref{4} gives the future development trend of this field from our point of view; Section \ref{5} concludes the entire paper.

\section{Method classification} \label{2}

\subsection{Development timeline} \label{2.1}

\begin{figure}[t]
    \centering
    \includegraphics[width=1.0\textwidth]{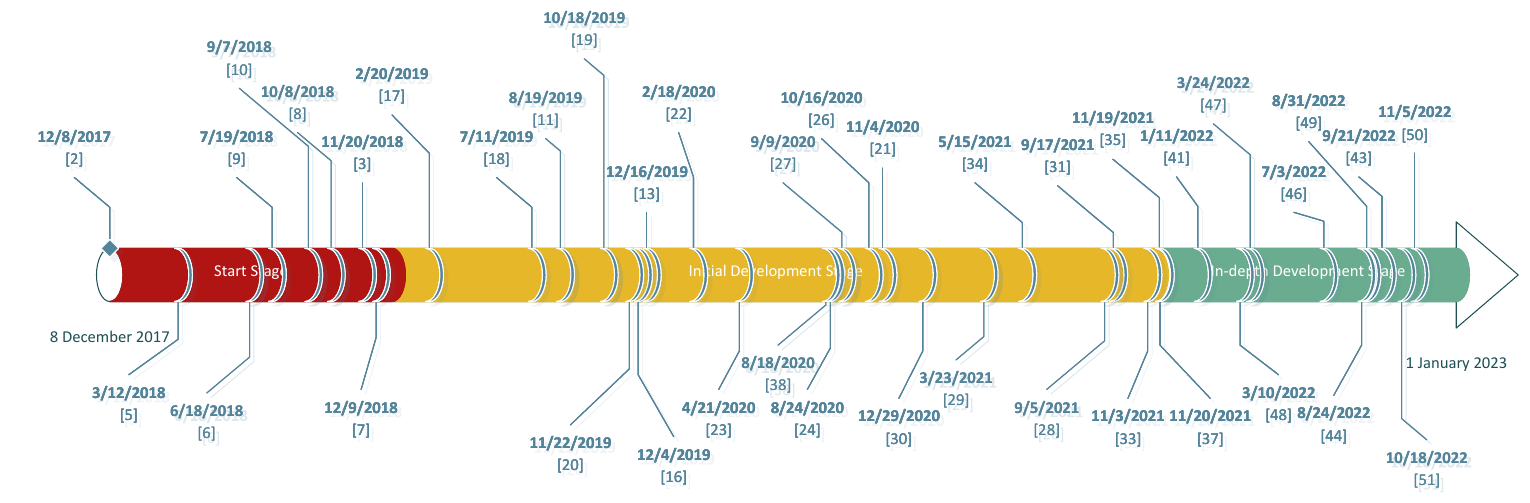}
    \caption{Development timeline of relevant research}
    \label{figure1}
\end{figure}

Although there have been many types of research related to webshell detection, methods based on AI technology have only gradually emerged since 2017, as shown in Figure \ref{figure1}.

Based on the timeline or technology development, we categorize the relevant research into the Start Stage, Initial Development Stage, and In-depth Development Stage. 

\subsubsection{Start Stage} \label{2.1.1}
In the Start Stage, researchers mainly focus on the preliminary exploration of AI-related algorithms in the field of webshell detection. Tian et al. \cite{RF2} used a simple Convolutional Neural Network (CNN) structure based on max-pooling and ReLU to detect webshells on HTTP traffic data. Building upon the work of Tian et al. \cite{RF2}, Zhang et al. \cite{RF3} proposed using character-level methods to transform webshell content features while preserving the sequential pattern features of the traffic content. They performed certain data cleaning tasks on the data flow, including URL decoding, BASE64 decoding, and binary data stream replacement, to eliminate the influence of encrypted traffic. In terms of the detector, they combined 2 neural network structures, CNN and Long Short Term Memory (LSTM) \cite{RF4}, to extract local key field features using CNN and text sequence features using LSTM. Fang et al. \cite{RF5} first combined the webshell static features (i.e. Longest string, Information entropy, Index of coincidence, etc.) with opcode sequences, and used FastText to train text classifier models. Similarly, Cui et al. \cite{RF6} combined static features with opcode sequences and designed a 2-layer classification model RF-GBDT. RF-GBDT initially uses Random Forest (RF) to obtain preliminary prediction results and then applies Gradient Boosting Decision Tree (GBDT) algorithm for further training and obtaining the final prediction results. Zhang et al. \cite{RF7} converted PHP script files into opcodes and applied N-gram to extract script features. They used Bagging strategy to integrate 5 individual classifiers (Support Vector Machine (SVM), K Nearest Neighbors (KNN), Naive Bayes (NB), Decision Tree (DT), CNN) to obtain final classification results. Yong et al. \cite{RF8} directly used Word Of Bag (WOG) method to vectorize script source code and fed it into a simple structured neural network for a binary classification task by Term Frequency-Inverse Document Frequency (TF-IDF) analysis. This method, which does not involve preprocessing, is easily affected by obfuscated webshell source code, making it difficult to guarantee the generalization ability of the method.

However, both traffic data and code files suffer from the problem of having excessively long input sequences. Unfortunately, AI-based algorithms, especially neural network models, are exceptionally sensitive to changes in input length. Therefore, Qi et al. \cite{RF9} employed downsampling methods to reduce information loss and computational cost, achieving higher detection accuracy through LSTM and pooling layer-based Deep Neural Network (DNN) structures. Specifically, after the vectorization process of generating token streams from the input using a lexical analyzer and converting the tokens into vector sequences through token embedding dictionaries, downsampling is used to convert the vector sequence into $k$ short streams for parallel computation and the LSTM-Pooling neural network structure is applied as a classifier. However, this method also suffers from the problem of too many artificially defined rules, such as constant replacement, which can lead to the loss of semantic information in the script source code. Liu et al. \cite{RF10} proposed the use of PL-CNN and Payload Classification-Recurrent Neural Network (PL-RNN) methods to detect network attack payloads and tested their performance on the DARPA dataset.

These Start Stage studies, despite their simple methods and many flaws and shortcomings, such as a small number of private datasets, unreasonable feature extraction, etc., have undoubtedly become pioneers in this field, laying the foundation for the emergence of numerous subsequent methods.

\subsubsection{Initial Development Stage} \label{2.1.2}

\begin{figure}[t]
    \centering
    \includegraphics[width=0.78\textwidth]{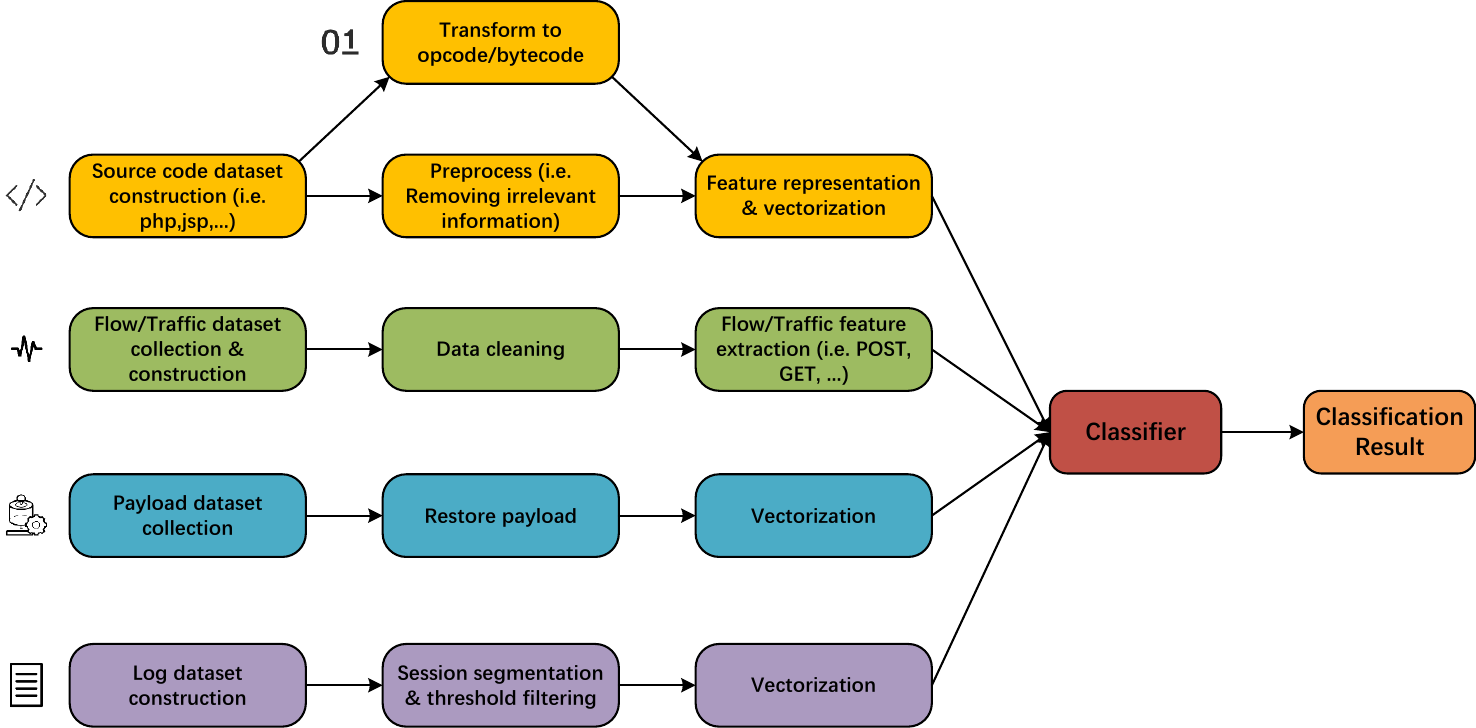}
    \caption{Detection workflow of different webshell datatypes}
    \label{figure2}
\end{figure}

Since 2019, research in this field has entered a stage of explosive development. A large number of AI-based methods have been applied to webshell detection research, and researchers have paid more attention to optimizing each stage of the complete detection pipeline, rather than just using simple neural network models as classifiers, as shown in Figure \ref{figure2}.

ShellBreaker \cite{RF11} summarizes 3 distinctly malicious features of webshells: the intensive usage of super global variables, adaption and automation, and data flow. It firstly parses PHP into Abstract Syntax Tree (AST), then performs syntactical analysis and semantic analysis. Wang et al. \cite{RF12} converted PHP source code into opcodes and subsequently used Bi-Gram to split the sample opcodes into opcode sequences. They calculated the word frequency matrix, performed word frequency analysis, selected the feature matrix of the training sample set using TF-IDF matrix, and set a frequency threshold (30\%) for filtering. Li et al. \cite{RF13} improved the neural network classifier by combining the attention mechanism with Gated Recurrent Unit (GRU) \cite{RF14} to detect and classify the samples that were vectorized using the Word2Vec \cite{RF15} word embedding algorithm. Nguyen et al. \cite{RF16} focused on using Yara rules for pattern matching to construct a black-and-white dataset. They performed de-duplication operations to obtain a higher-quality training set for enhancing the training effect of the model. Lv et al. \cite{RF17} comparatively tested the performance impact of different word vectorization methods (e.g. Improved One-hot, Bag-of-words, Word2Vec) on CNN classifier. Tao et al. \cite{RF18} optimized the preprocessing steps by first recovering the malicious payload before vectorization and enhancing the algorithm security. 

Tianmin et al. \cite{RF19} proposed a composite sample representation method using opcode+N-Gram+TF-IDF for code vectorization. They compared and tested 4 different machine learning algorithms, including eXtreme Gradient Boosting (XGBoost), Multi-Layer Perceptron (MLP), RF, and NB. Wu et al. \cite{RF20} took an alternative approach to detect webshell communication with exact sessions derived from weblogs. They employed IP fields and user agent fields to roughly divide the collected logs into different sessions, calculated the time intervals in each session, set thresholds to identify sessions in more detail, and identified webshells through the final hidden Markov Chain model and LSTM. This approach undoubtedly provides new insights for webshell detection. RF-AdaCost \cite{RF21} enhances RF-GBDT \cite{RF6} by merging static features (e.g. Information entropy and Coincidence index) of PHP source code with opcode sequences to improve the detection efficiency of webshell. However, this method still has many defects, for example, the length of the longest word in the text is judged in its set of feature filtering rules, but the fact is that even white samples may contain long strings. WSLD \cite{RF22}, on the other hand, combines fuzzy matching with RNN for the first time to propose a heuristic detection algorithm for webshell. This method is no longer limited by the script language used to create webshells, but divides the essence of webshells into 2 parts: data transmission and data execution. Based on this idea, webshell attacks can be categorized into the following 4 categories:

\begin{enumerate}[label=\roman*]
    \item Modifications in commands and functions based on data execution and data transmission;
    \item Confusion variations based on data execution and data transmission;
    \item A variation based on web server data execution and data transmission configuration;
    \item Data execution and data transmission deformation methods based on language characteristics.
\end{enumerate}

In this way, the authors can utilize webshell itself to construct a fuzzy hash sample library, define the main rules of the feature library based on webshell data execution and data transmission at a fine-grained level. Statistical features of webshells are then employed to construct an RF-based fuzzy matching algorithm. WS-LSMR \cite{RF23} optimizes the ensemble learning approach for webshell detection. The authors concluded that ensemble learning for webshell detection suffers from the following difficulties:

\begin{itemize}
    \item How to select the optimal feature subset of malicious webshell scripts;
    \item How to select base learner in ensemble learning;
    \item How to calculate the weight parameters of the base learner in ensemble learning.
\end{itemize}

Therefore, the authors extracted a feature vocabulary based on 1-gram and 4-gram opcode, and applied feature selection using 1-gram and 4-gram algorithm. They constructed a functional vocabulary for the simultaneous detection of encrypted/ unencrypted webshell files and integrated Logistic Regression (LR), SVM, MLP, and RF individual classifiers with weighted values. However, this method also exposes the drawback of slow detection time in the ensemble learning scheme. Additionally, the authors pointed out that webshell detection based on static features has the problem of being prone to false positives and cannot detect new unknown webshells. The problem with webshell detection based on dynamic features is the need to maintain a large behavior characteristics library. Therefore, in their subsequent work \cite{RF24}, they combined both static and dynamic features as algorithmic features to construct a comprehensive feature set. The authors further pointed out that there are 3 main difficulties in webshell detection:

\begin{enumerate}[label=\roman*]
    \item Extremely unbalanced datasets;
    \item Irrelevant or redundant features;
    \item Certain limitations in the detection algorithm.
\end{enumerate}

Furthermore, they also performed a de-duplication operation in the pre-processing stage and applied the genetic algorithm to extract the validity of feature dimensions. They oversampled the dataset based on SMOTE algorithm \cite{RF25} to mitigate the impact of data imbalance on classifiers and prevent the explosion of feature quantity. Additionally, the authors introduced ensemble learning algorithms with K-fold cross-validation. This approach was also adopted by Jinping et al. \cite{RF26}. However, such methods still require manual large-scale feature engineering, and static feature extraction also relies on the identification of extremely long strings, resulting in weak applicability and generalization capabilities.

Webshell malicious feature fusion is also the focus of relevant research at this stage. Zhu et al. \cite{RF27} proposed a multi-view feature fusion method for classifying PHP webshells. They extracted lexical features, syntactic features, and abstract features, and used Fisher to rank the importance of these features, with SVM as the classifier. Huang et al. proposed the UTANSA framework \cite{RF28}, which unifies text features and AST node features of 2 script languages, PHP and JS. However, this method still uses the traditional TF-IDF algorithm to obtain vectorized features and employs the RF algorithm as the classifier. Pan et al. \cite{RF29} extracted partial node features from the PHP AST structure and combined textual features with static features to construct the feature matrix. They employed supervised learning for webshell classification tasks.

Zhang et al. \cite{RF30}, on the other hand, focused on comparing the representation schemes for webshell script source code. They tested webshells in both PHP and JSP script languages. For PHP, they converted it to AST structure, while JSP was processed as bytecode sequences. Experimental results indicate that AST is a better representation scheme than bytecode. Because bytecode only contains instruction information, but ignores valid information such as parameters, while AST can completely react to the semantic information of script source code, and is more suitable for the 2-gram approach. In addition, they also took into account the problem of detection rate and detection bandwidth. Wu et al. \cite{RF31} innovatively used a CNN-based webshell detection method improved by reinforcement learning. They argued that the key to improving classification accuracy lies in appropriately combining the advantages of neural network automatic feature selection and expert knowledge-based methods. Specifically, they utilized Asynchronous Advantage Actor-Critic (A3C) \cite{RF32} reinforcement learning for automated feature selection and maximized the expected accuracy of the CNN classifier on the validation dataset through sequential interaction with the feature space. 

Building upon the work of Nguyen et al. \cite{RF16}, Le et al. \cite{RF33} proposed a hybrid webshell detection method based on a combination of pattern matching and deep learning. They employed yara-based pattern matching on a clean dataset, used CNNs to train predictions on webshells with OCI vectors (opcodes) converted from .NET code, and employed YBPM on a set of files predicted to be benign by a deep learning model to check for the presence of false positives. However, the study has a limited reference value due to the small amount of experimental data. Zhou et al. \cite{RF34} applied LSTM as a classifier after converting PHP to opcode and using Word2vec to perform word vectorization operations. They obtained an interesting finding through experiments that increasing the number of layers in the LSTM neural network did not necessarily improve accuracy. Instead, a single-layer LSTM model achieved the highest accuracy. Cheng et al. \cite{RF35} proposed a webshell detection model based on the Text-CNN \cite{RF36} model. They extracted PHP and JSP scripts into AST structures and further transformed them into AST node text sequences, using Text-CNN as the webshell classifier. Hannousse et al. \cite{RF37} developed $RF-DNN^2$, which integrates 2 deep learning models. The first neural network model receives vectorized source code as inputs, while the second one receives vectorized opcodes. The final classification result is obtained through the RF classifier. However, this method still needs to manually define and extract syntactic features, lexical features, and static features for a total of 100 features, which does not fully utilize the advantages of feature extraction of deep learning models. Yong et al. \cite{RF38} further improved the ensemble learning approach. They employed Extremely Randomized Trees (ET) as individual classifiers and used Heavyweight WDS (HWDS) for ensemble integration.

During the Initial Development Stage, relevant studies have made varying degrees of exploration and attempts in the entire detection process, achieving certain progress. However, these methods still have more or less deficiencies in the fields of detection dataset, universality of methods, extraction of features, etc. Theoretical innovation remains relatively scarce.

\subsubsection{In-depth Development Stage} \label{2.1.3}

Since the end of 2021, with the rise of various variants of BERT \cite{RF39} models (e.g. CodeBERT \cite{RF40}) and new neural network algorithms, webshell detection methods based on AI technology have entered the In-depth Development Stage. Simple individual classifiers or machine learning algorithms are no longer common, and related research has basically penetrated into the theoretical process level of modeling methods. Xie et al. \cite{RF41} used the PHP extension of Vulcan Logic Disassembler (VLD) to obtain the opcode sequence of PHP files, and converted the opcode to a fixed 100-dimensional vector with the help of Word2Vec. Subsequently, they performed feature extraction using EDRN neural network \cite{RF42} and obtained the final classification results through the $Sigmoid$ function. Le et al. \cite{RF43} open-sourced their detection scheme in traffic data based on previous work \cite{RF16}. This rule-based detection uses filters to determine HTTP traffic and extracts a total of 79 feature representations. They designed a DNN model for deep analysis to detect webshell traffic from benign traffic. They used batch normalization as regularization to help reduce overfitting and replaced the dropout scheme to prevent the deep neural network model from losing too much information during computation. However, this method still cannot completely overcome the limitations of feature engineering. 

An et al. \cite{RF44} focused their research on addressing the problem of long PHP script file selection. They argued that several key issues to be addressed by current PHP webshell detection schemes lie in the challenges of text selection when coping with long PHP scripts, lexical ambiguity in programming language, and the challenge of decline of generalization ability to unseen PHPs if training samples are not treated reasonably. Whereas traditional methods and machine learning approaches have difficulty in recognizing webshells that do not have signatures or patterns in the database (i.e. 0-day webshells). To address this, they proposed a novel two-stage webshell detection framework. In Stage 1, a TextRank-based \cite{RF45} sentence-level text selection model is used to preserve code semantics via extracting high-value code lines. Previous text filter algorithms, such as token filter method, belong to a micro lexical-level granularity which might ignore the semantics in the neighboring context. On the other hand, the text truncation method belongs to macro document-level granularity which might lose the focus of high-value information of the text. In Stage 2, instead of using the Word2Vec word embedding algorithm, they utilized a CodeBERT-based token embedding model to resolve lexical ambiguity and generate token representation vectors. The advantage of this approach is that CodeBERT goes through a pre-training stage of the code and the representation vectors of the entire selected lines of code are represented by the output vectors at the $[CLS]$ position. Since the $[CLS]$ vector contains the synthesized semantic information of the PHP script, it can be directly applied to webshell detection tasks in combination with lightweight classifiers such as $softmax$. Furthermore, downstream deep learning classifiers can also be used to label the vectors and extract more semantic information. The authors also paid additional attention to data privacy protection. They argued that mixing the collected benign samples with webshell samples into a complete dataset and selecting training and validation sets can lead to data leakage because benign PHP scripts in the same GitHub project may have project-specific code expressions, such as the same class name, the same function name with the same programming style. Whereas, using a randomly selected data segmentation method will split a part of the benign project in the training dataset and another part in the validation set, which in turn leads to a high degree of similarity between the validation set and the training set. The authors ultimately chose Text-CNN \cite{RF36} as the downstream classifier. However, this approach lacks convincing performance comparisons and testing against other related research, and the effectiveness of the result is unconvincing. 

Not coincidentally, MSDetector \cite{RF46} also recognizes the excellent performance of CodeBERT in code-related tasks. Since the CodeBERT model itself does not introduce webshell-related script source code in the pre-training stage, MSDetector pre-trains CodeBERT on the AST Node Tagging (ANT) tasks and designs a novel script sequence representation scheme that combines text sequences and lexical token sequences to perform webshell classification tasks with subsequent classifiers. However, MSDetector still relies on manual pre-filtering in the pre-processing stage, which requires up to 1 month of pre-filtering time with less than 10000 data volumes, making it difficult to migrate to the webshell detection tasks in the context of big data. Additionally, language models like CodeBERT that contain a massive number of parameters require massive data support in the pre-training stage, and the insufficient amount of data in MSDetector makes it difficult to guarantee the effectiveness of pre-training. 

Liu et al. \cite{RF47} proposed a webshell detection method based on bi-directional GRU and attention mechanism that supports multiple languages (i.e., PHP, JSP, JAVA, etc.). This method can directly extract abstract features of webshells without relying on feature engineering. Gogoi et al. \cite{RF48}, while not making significant innovations in the detection process, analyzed common function calls and super global variable calls in PHP webshells using LSTM neural network. 

Pu et al. \cite{RF49} focused their attention on webshells based on JSP script language. They used Tomcat Server to convert JSP files into servlets in the form of Java classes to generate bytecode. They applied BERT to generate word embeddings that were fed into a classifier consisting of XGBoost/Bi-LSTM for identification. Jiang et al. \cite{RF50} examined multimodal webshell detection schemes. They comprehensively analyzed webshells through multiple dimensions (i.e. traffic, logs, page associations, etc.), and utilized TinyBERT to generate feature vectors. However, the pre-processing stage of this method involves overly strict rules (e.g. Only the contents of GET and POST fields are considered to be of the highest value in the traffic packets), leading to the loss of a significant amount of critical information, making it difficult to perform feature association analysis. Based on previous work \cite{RF37}, Hannousse et al \cite{RF51}. proposed a webshell detection scheme that can simultaneously detect webshells in four languages including PHP, JSP, ASP, and ASPX at the source code level. They analyzed webshell attacks and obfuscation methods in detail (e.g. attack methods: command execution, file upload; obfuscation methods: letter slicing, code encryption, code obfuscation) and proposed a fine-grained data de-duplication scheme (e.g. filtering through $MD5$ hash and secondary de-duplication after disassembling opcode by VLD). Experimental data shows that the detection accuracy at the source code level is higher than opcode, since two PHP scripts with completely different code logic and semantic logic may be identical after converting to opcodes. Therefore, relying solely on opcodes for black-white sample classification would have limitations and performance disadvantages.

\subsection{Model categories} \label{2.2}

We divide the relevant research into categories according to machine learning-based methods, deep learning-based methods, and hybrid models (including but not limited to using ensemble learning to combine machine learning and deep learning methods, comparative testing of machine learning and deep learning methods, etc.). The summary is shown in Table \ref{Table1}.

\begin{table}[H]\footnotesize
    \centering
    \caption{Classification of model categories}
    \begin{tabularx}{\linewidth}{cXX}
    \toprule
    Order & Method category \centering & References \\
    \midrule
    1 & Machine learning \centering & \cite{RF21} \cite{RF6} \cite{RF5} \cite{RF27} \cite{RF11} \cite{RF28} \\
    \hline
    2 & Deep learning \centering & \cite{RF2} \cite{RF3} \cite{RF8} \cite{RF9} \cite{RF12} \cite{RF13} \cite{RF16} \cite{RF17} \cite{RF18} \cite{RF10} \cite{RF31} \cite{RF33} \cite{RF34} \cite{RF35} \cite{RF41} \cite{RF43} \cite{RF44} \cite{RF46} \cite{RF47} \cite{RF48} \cite{RF50} \\
    \hline
    3 & Hybrid models \centering & \cite{RF7} \cite{RF19} \cite{RF20} \cite{RF22} \cite{RF23} \cite{RF24} \cite{RF26} \cite{RF30} \cite{RF37} \cite{RF38} \cite{RF49} \cite{RF51} \cite{RF29} \\
    \bottomrule
    \end{tabularx}
    \label{Table1}
\end{table}

As can be seen from Table \ref{Table1}, deep learning-based methods occupy the vast majority of related studies, while methods using only machine learning algorithms are very limited. Therefore, it can be predicted that deep learning-based methods and hybrid models will be the future research trends in this field.

\subsection{Detection datatypes} \label{2.3}

In the pipeline shown in Figure \ref{figure2}, we divide the data into 6 categories based on different datatypes: Source code data, Flow/Traffic data, Payload data, Log data, Opcode/Bytecode, and others. Similarly, we categorize the related studies based on datatypes, as summarized in Table \ref{Table2}.

\begin{table}[H]\footnotesize
    \centering
    \caption{Method classification based on detection datatypes}
    \begin{tabularx}{\linewidth}{cXX}
    \toprule
    Order & Datatype \centering & References \\
    \midrule
    1 & Source code data \centering & \cite{RF9} \cite{RF13} \cite{RF17} \cite{RF30} \cite{RF35} \cite{RF37} \cite{RF44} \cite{RF46} \cite{RF47} \cite{RF48} \cite{RF51} \cite{RF27} \cite{RF11} \cite{RF28} \\
    \hline
    2 & Flow/Traffic data \centering & \cite{RF2} \cite{RF3} \cite{RF31} \cite{RF43} \\
    \hline
    3 & Payload data \centering & \cite{RF18} \cite{RF10} \\
    \hline
    4 & Log data \centering & \cite{RF20} \\
    \hline
    5 & Opcode/Bytecode \centering & \cite{RF7} \cite{RF8} \cite{RF12} \cite{RF16} \cite{RF19} \cite{RF21} \cite{RF23} \cite{RF24} \cite{RF26} \cite{RF30} \cite{RF33} \cite{RF34} \cite{RF37} \cite{RF41} \cite{RF38} \cite{RF49} \cite{RF6} \cite{RF5} \cite{RF29} \\
    \hline
    6 & Others \centering & \cite{RF22} \cite{RF50} \\
    \bottomrule
    \end{tabularx}
    \label{Table2}
\end{table}

As can be seen from Table \ref{Table2}, it can be observed that in the relevant research, there is a higher number of studies focusing on Source code data and Opcode/Bytecode, while the remaining datatypes are relatively few. However, different datatypes have their advantages and limitations, which we will explore in depth in Section \ref{3}.

\subsection{Summary} \label{2.4}

\begin{figure}[H]
    \centering
    \includegraphics[width=0.66\textwidth]{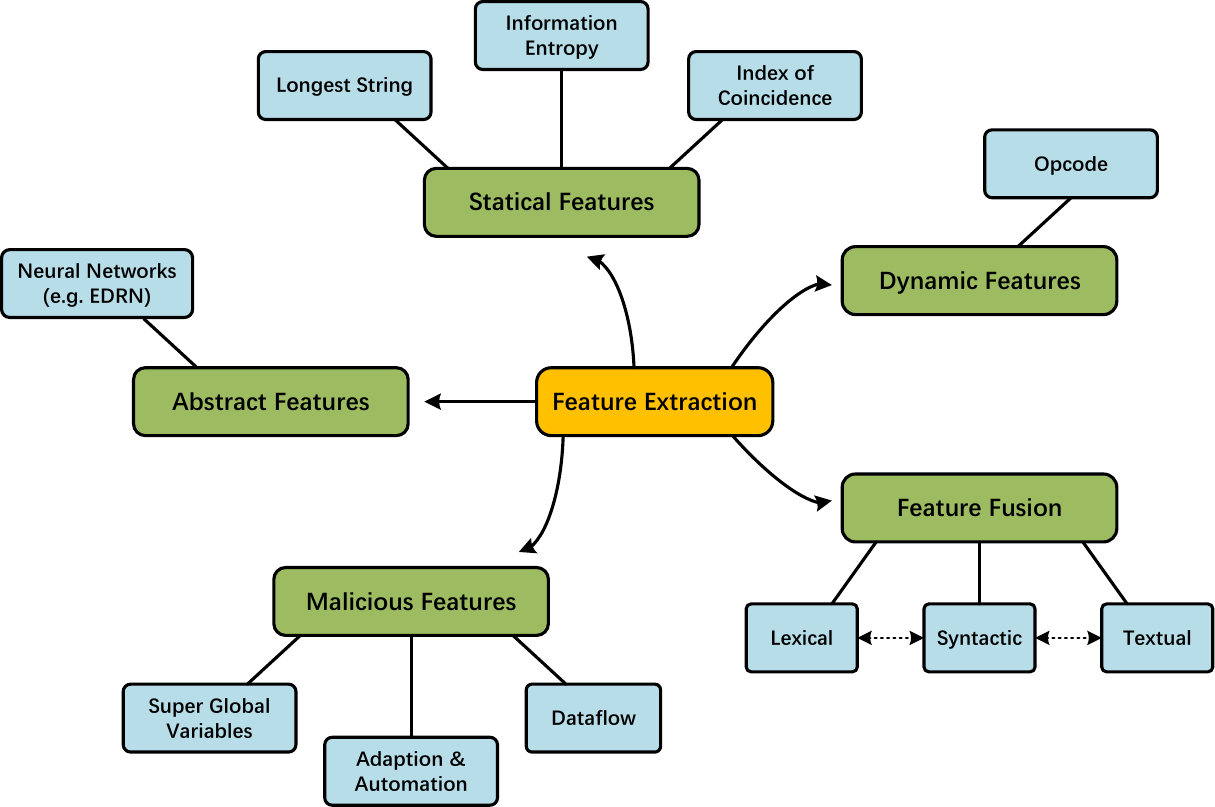}
    \caption{Summary of feature extraction methods in related studies}
    \label{figure3}
\end{figure}

\begin{figure}[H]
    \centering
    \includegraphics[width=0.66\textwidth]{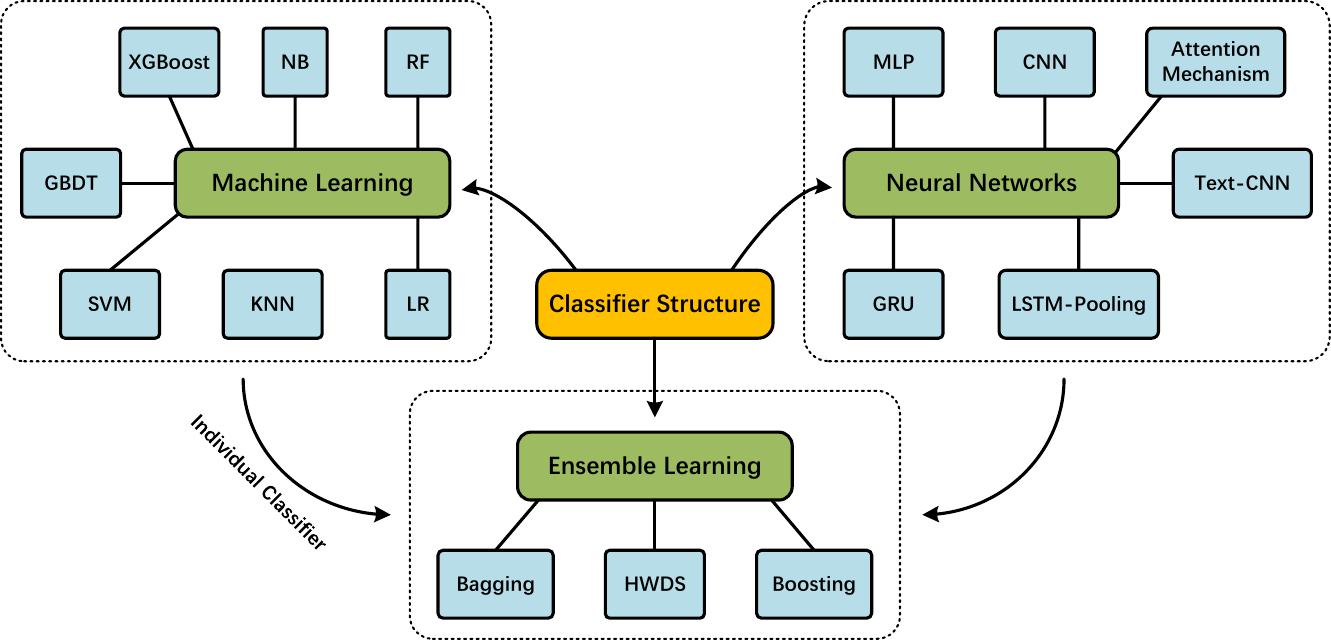}
    \caption{Summary of classifier structures in related studies}
    \label{figure4}
\end{figure}

\begin{figure}[H]
    \centering
    \includegraphics[width=0.38\textwidth]{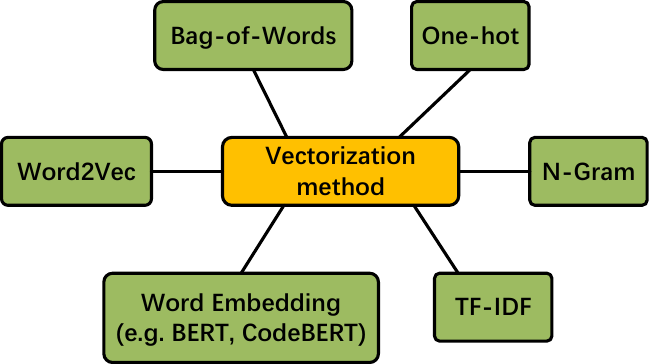}
    \caption{Summary of vectorization methods in related studies}
    \label{figure5}
\end{figure}

We summarize the feature extraction, classifier structures, and vectorization approaches of relevant research respectively, as shown in Figure \ref{figure3}, Figure \ref{figure4}, and Figure \ref{figure5}.

Figure \ref{figure3} indicates that most studies focus on extracting statical features, dynamic features, and malicious features from webshells. However, these features contain many biases of human-defined rules, which cannot fully leverage the adaptive feature selection and extraction capabilities of AI algorithms. In the In-depth Development Stage, researchers gradually shift to abstract feature extraction and feature fusion, such as EDRN neural networks. It is therefore foreseeable that more researchers will delve into the field of abstract feature extraction and feature fusion in the future.

Figure \ref{figure4} reflects that the classifier structures used in existing methods can be divided into 3 major categories: machine learning, neural networks, and ensemble learning. Methods during the Start Stage and Initial Development Stage mostly employ machine learning classifiers, which have poor generalization ability and classification stability despite the simplicity of the structure. Subsequent studies have gradually shifted towards neural networks and ensemble learning methods, but they also suffer from the drawbacks of low efficiency and excessive resource consumption, which we will further discuss in Section \ref{3}.

In terms of the vectorization approaches, early methods use algorithms such as One-hot, N-Gram, Bag-of-Words, TF-IDF, etc., which have low information density and lack contextual association. Subsequent studies have begun to adopt the Word2Vec algorithm and word embedding schemes such as BERT, which can effectively incorporate sequential structure and contextual information, and improve the integration of the system.

\section{Critical issues and challenges} \label{3}

\subsection{Appropriate data representation} \label{3.1}
As described in Section \ref{2.3}, we categorize the detection datatypes in relevant research. However, researchers hold different views on which data representation scheme is suitable for webshell data. Hannousse et al. \cite{RF51} argued that the detection accuracy of source code is higher than opcode. This is because two PHP scripts with completely different logic and functionality at the source code level may appear identical after being converted to opcode. Similarly, Zhang et al. \cite{RF30} also believed that the representation of source code is superior to opcode because the process of generating opcode involves the line-by-line conversion of source code statements, which only considers instructions and ignores the corresponding opcode parameters, resulting in a lack of valuable information. However, Kang et al. \cite{RF21} suggested that combining static features with opcode can improve the efficiency of webshell detection. Similarly, $RF-DNN^2$ \cite{RF37} believed that using opcode detection can avoid explicit de-obfuscation and effectively detect potential webshell variants.

In this paper, we firmly believe that source code undoubtedly contains more webshell features than opcode. Since opcode is essentially compiled from source code using tools like VLD, it inevitably loses critical information such as source code parameters. In fact, any manually defined processing rules or procedures will inevitably result in the loss of some webshell source code information. From the perspective of research timeline order, most of the research related to webshell detection focusing on opcode is concentrated in the Start Stage and Initial Development Stage. This is because converting source code to opcode greatly simplifies subsequent processing steps (i.e. de-obfuscation steps) and facilitates feature extraction due to its relatively uniform format. However, as research progresses, the deficiencies of opcode in losing a significant amount of semantic information become increasingly evident. Therefore, novel data representation methods such as AST can be considered.

However, all source code-based detection schemes must address the issue of cross-language and model generalization capabilities. Different script languages, such as PHP and JSP, have significant differences in the generated AST structures. To convert them into a unified format suitable for subsequent processing, custom token conversion rules need to be defined. In contrast, webshell detection solutions based on flow/traffic/ payload/logs do not rely on source code language, thus overcoming the limitations of source code analysis. However, for data representation schemes based on flow/traffic/payload, it also needs to carry out complex steps such as shelling, decryption, and payload recovery to carry out the subsequent feature extraction and classification process. This is because most of the network traffic data in today’s network is shelled/encrypted, and webshells employ even more sophisticated techniques to subtly hide and disguise themselves.

Therefore, no matter what kind of data representation is chosen, careful design of pre-processing steps and procedures is necessary to retain webshell features as much as possible, eliminate interference from redundant information, and enhance the efficiency and accuracy of webshell detection.

\subsection{Machine learning VS deep learning VS hybrid model} \label{3.2}
From the summary in Section \ref{2.2}, it can be seen that the number of relevant studies using deep learning-based methods is much higher than that of machine learning-based algorithms. The advantage of deep learning-based methods over machine learning-based algorithms depends on their ability to directly extract abstract features of webshells by their non-linear computational characteristics, whereas machine learning-based algorithms require manual feature engineering. However, any manually defined feature engineering method is inevitably limited by cognitive limitations and biases, resulting in the loss of webshell features. For example, RF-AdaCost \cite{RF21} uses the maximum length of tokens in text as one of the features for determining webshells, which is obviously an unreasonable cognitive bias because long string tokens can also exist in benign samples. However, research on webshell detection using deep learning-based methods also faces the problem of input length. This problem is prevalent in various data representation categories such as source code, AST, opcode, and traffic data. Since the introduction of the Transformer model \cite{RF52} in 2017, attention mechanism has been widely used in various deep learning models, such as BERT \cite{RF39}, TransformerXL \cite{RF53}, RoBERTa \cite{RF54}, and the recent popular models like GPT-4 \cite{RF55} and LLaMA \cite{RF56} for its excellent feature extraction capability and parallel computation properties. However, these attention-based models also have limitations on input length. Taking the BERT model as an example, its maximum input is limited to 512 tokens, and the performance overhead required for model training grows dramatically as the input length increases. Although models like Longformer \cite{RF57} can handle longer inputs, they are still insufficient when dealing with large webshell code exceeding 20000 tokens in length. 

\begin{figure*}[t]
    \centering
    \includegraphics[width=0.82\textwidth]{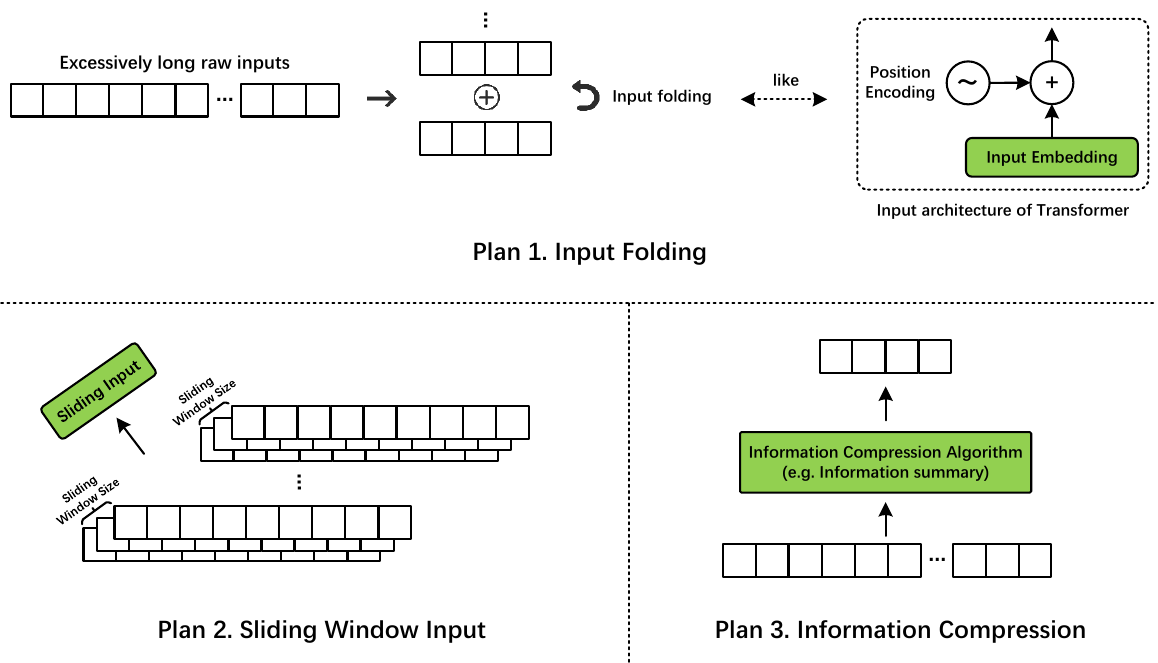}
    \caption{Feasible ideas to address the problem of excessively long inputs}
    \label{figure6}
\end{figure*}

To address the issue of input length, various approaches have been used in related studies. Most studies in the Start Stage and Initial Development Stage directly truncate excessively long webshells, which results in the loss of a significant amount of critical information. Studies in the In-depth Development Stage provide some new solutions but also have several drawbacks. The TextRank model used by An et al. \cite{RF44} forgets the contextual relevance between statements although it can take into account the semantics in neighboring contexts. MSDetector \cite{RF46} uses a custom lightweight AST representation method that effectively compresses the length but also loses some semantic information of webshells. Researchers need to design more reasonable feature extraction methods to deal with excessively long webshells while minimizing feature loss. For example, Vanilla Transformer can preserve the original input information by adding Position Encoding to the original input, while also incorporating contextual position relationships into the fused representation. So, for extremely long webshells, can we use input folding to obtain the fused input representation? Or can we use a sliding window approach to read long inputs in batches while retaining the contextual information? Or designing new information compression algorithms to preserve the core malicious features of webshells? These may be some feasible ideas for addressing long webshell inputs and require further exploration and investigation by researchers, as shown in Figure \ref{figure6}.

For hybrid models, ensemble learning can provide a slight performance improvement compared to a single classifier, but it also significantly increases the performance overhead of training and running the system. In addition, compared to the choice of downstream classifiers, the processing flow of upstream webshell scripts is the key to performance improvement. 

\subsection{Data balance VS data imbalance} \label{3.3}
Related studies hold different views on whether the dataset should be balanced or not \cite{RF58}. Cheng et al. \cite{RF35} used an extremely imbalanced dataset in their experiments, with a ratio of $1:5.25$ between black and white samples. On the other hand, Jinping et al. \cite{RF26} specifically employed sample balancing algorithms such as over-sampling and down-sampling to alleviate the problem of data imbalance. Imbalanced datasets are consistent with the real-world situation of webshells, where they represent only a tiny fraction of all network-scripts. However, using such dataset distribution proportions in experiments is not conducive to assessing the effectiveness of detection algorithms because even simply labeling all detection samples as white samples can achieve a high "accuracy". We believe that in the experimental environment, it is important to maintain a balanced ratio of black-and-white datasets to more realistically and effectively evaluate the actual performance of a detection algorithm.

\subsection{Dataset problems} \label{3.4}
In the survey conducted by Abdelhakim et al. \cite{RF1}, it is found that the current webshell detection research lacks a benchmark dataset. Although they have released a webshell dataset containing black-and-white samples in PHP and JSP scripting languages, the number of webshell scripts included is very limited. The quantity of JSP webshells is even less than 1000, which is insufficient to support the training of neural network models with a large amount of data. In fact, almost all the related studies summarized in this paper have this problem, with most model methods having a training dataset size of less than 10000. Such a small training dataset may be sufficient for some traditional machine learning algorithms (i.e. SVM, RF, DT, etc.), but it is far from enough to support the training of neural network models, especially those known for their massive parameters like BERT and emerging Large Language Models (LLMs). Even if successful training is achieved, it will lead to severe overfitting issues, resulting in weak generalization and limited applicability of the methods. For example, MSDetector \cite{RF46} performs pre-training tasks on CodeBERT, but with such a small amount of data, it is difficult to ensure the effectiveness of pre-training. Through our actual investigation, it is evident that the webshell datasets available from sources like GitHub and Mendeley are indeed very limited. One reason is the sensitive and compliant nature of the content in webshell datasets. To overcome this challenge, we suggest that researchers consider the following aspects for generating and expanding data samples:

\begin{enumerate}
    \item Cooperate with internationally renowned internet enterprises and security companies to construct the dataset jointly. Large internet companies are subject to countless cyber-attacks every year. Although such data is often not made available to the public due to privacy protection principles, researchers can greatly improve the plight of insufficient data by reaching cooperation with relevant companies;
    \item Expand sample data with the help of Artificial Intelligence for Generating Content (AIGC) techniques. For example, using methods like Generative Adversarial Networks (GAN) \cite{RF59} for data sample generation, and applying reinforcement learning techniques \cite{RF60} for data augmentation \cite{RF61}. Although GAN networks are primarily used for generating image data \cite{RF62} and serialized data \cite{RF63}, researchers can transfer them to the task of generating webshell scripts through appropriate transfer learning tasks \cite{RF64}. Regarding data augmentation methods, current research mainly focuses on mainstream AI fields such as Natural Language Processing (NLP) \cite{RF65}, time series data \cite{RF66} and Computer Vision (CV) \cite{RF67}. However, considering that data augmentation has been successfully applied to code-related tasks such as code generation \cite{RF68}, code readability classification \cite{RF69}, it can naturally be extended to webshell data. It should be noted that both data sample generation methods and data augmentation methods should consider the validity of the newly generated webshells to ensure that the model can still learn real and effective webshell features in the expanded dataset, to improve the model's classification accuracy and generalization ability.
\end{enumerate}

\section{Development trends} \label{4}

\subsection{Few-shot learning} \label{4.1}

Due to the complex feature of webshell scripts, a limited training set is insufficient to cover all types of webshell scripts. Therefore, an excellent webshell detection method with high practical value should possess strong few-shot learning \cite{RF70} capabilities rather than merely achieve outstanding performance on its private dataset. Mainstream few-shot learning methods, such as MAML meta-learning \cite{RF71}, have been successfully applied to the task of anomaly detection on multivariate time series data \cite{RF72}. However, none of the existing studies related to webshell detection have considered few-shot learning strategies. Therefore, few-shot learning can be regarded as one of the important future directions for webshell detection research.

\subsection{Federated learning} \label{4.2}

From Section \ref{2.2} on relevant classifications, it can be observed that the use of deep learning and hybrid models holds an absolute advantage. However, with the iterative development of deep learning techniques and the improvement of model inference capabilities, the demand for computing power has also significantly increased. This has been illustrated in the research conducted by An et al. \cite{RF44} and Cheng et al. \cite{RF46}. In cases where individual GPU computational resources are limited, the distributed training strategy \cite{RF73} of federated learning \cite{RF74} can effectively alleviate the computational pressure, but it is also necessary to consider the data privacy protection issue of communication between different servers. However, it is foreseeable that with the application of more advanced deep learning algorithms, federated learning strategies will become an important support for model training.

\subsection{Continual learning and life-long learning} \label{4.3}

Webshell attacks and detection methods are engaged in a game of upward, constantly evolving stage. When attackers develop new webshell scripts for network attacks, internet vendors tend to react and make decisions within a short period (usually days or weeks), repairing and improving the system in time through patching. Subsequently, attackers seek new attack methods and develop new webshell attack scripts. Therefore, continual learning and life-long learning \cite{RF75} are crucial for webshell detection methods. Life-long learning not only enables webshell detection solutions to go beyond learning only the webshell features contained in the current training set but also allows for continuous supplementation of new webshell features without re-training the model, as well as to compensate for the limitations of deep learning-based detection paradigms when facing 0-day webshells. Furthermore, life-long learning enables the model to retain the detection capability for the original task while adjusting and expanding the model for more complex tasks and to improve old knowledge and storage of new knowledge for future use.

\subsection{Large Language Models} \label{4.4}

Since the birth of LLMs \cite{RF76}, they have made an indelible impact in various fields such as chat conversations, image generation, etc. There are numerous LLM models in different subfields with different focuses. For example, GLM\cite{RF77} and GLM2 models tend to prioritize open-source and lightweight to meet the deployment needs of personal terminals. DALLE \cite{RF78} focuses on AI image generation, while FATE-LLM \cite{RF79} is biased towards application scenarios under the federated learning paradigm. LLM’s contextual reasoning and semantic understanding capabilities have been significantly ahead of early neural networks like LSTM and GRU. LLMs have been successfully applied to several subfields of code security, including vulnerability exploit code generation \cite{RF80, RF81}, automated penetration testing development \cite{RF82}, vulnerability description mapping \cite{RF83}, program repair \cite{RF84, RF85, RF86}, bug reproduction \cite{RF87}, vulnerability detection \cite{RF88} and LLM fuzz tuning \cite{RF89}. Through reasonable Prompt design \cite{RF90} and model fine-tuning \cite{RF91}, LLMs can even analyze the core functions of webshell source code layer by layer. Therefore, giving full play to the code reasoning ability of LLM, applying them effectively in webshell detection research, as well as constructing webshell detection infrastructure based on LLM technology will become one of the important application scenarios of LLM in the field of code security.

\subsection{New methodological paradigms} \label{4.5}

There are still many details that need to be improved and considered in the complete webshell detection pipeline. For example, since the introduction of the BERT model, pre-training is an effective way to improve the model's performance on downstream tasks. However, only MSDetector \cite{RF46} has made a preliminary attempt at pre-training tasks specifically for webshells, but the effectiveness is limited due to the small amount of pre-training data. Additionally, few methods have taken into account the detection rate (for source code/opcode-type data) and bandwidth (for traffic data). However, in practical applications, efficiency and time complexity are crucial aspects to consider, especially when facing high-volume large-scale detection scenarios. For instance, lightweight neural network models like Informer \cite{RF92} can be used. Moreover, novel neural network architectures such as Graph Neural Networks \cite{RF93} (GNN), can also be applied in the task of extracting feature associations between different webshells, to further optimize the detection performance of the methods. These novel methods, ideas, and strategies will become important development trends in future webshell research.

\section{Conclusion} \label{5}

This paper provides an overview of webshell detection methods based on AI techniques. We categorize related research according to the timeline and detection methods, and provide detailed explanations of the core key techniques in different studies, pointing out their limitations. We further analyze and discuss the key issues and future development directions in this research field. We hope that this review will help readers gain a comprehensive understanding of the research on AI technology in the field of webshell detection. In conclusion, AI-based webshell detection methods are still in the early stages, and we are confident that more innovative techniques and frontier theories based on AI techniques will emerge in the future. Correspondingly, detection techniques will also evolve into new forms.

\section*{Acknowledgement}

The authors thank the anonymous reviewers for their insightful suggestions on this work.
%% The Appendices part is started with the command \appendix;
%% appendix sections are then done as normal sections
%% \appendix

%% \section{}
%% \label{}

%% If you have bibdatabase file and want bibtex to generate the
%% bibitems, please use
%%
\bibliographystyle{elsarticle-num}

\bibliography{main}

%% else use the following coding to input the bibitems directly in the
%% TeX file.

%%\begin{thebibliography}{00}

%% \bibitem{label}
%% Text of bibliographic item

%%\bibitem{}

%%\end{thebibliography}
\end{document}